\documentclass[showpacs,aps]{revtex4}
\usepackage{amsmath}
\usepackage{times}
\usepackage{epsfig}
\usepackage{amssymb}
\usepackage{graphicx}
\usepackage{pstricks}
\usepackage{psfrag}

\begin{document}
\today
\title{
Current Fluctuations in One Dimensional Diffusive Systems 
with a Step Initial Density Profile
}
 \author{Bernard Derrida and Antoine Gerschenfeld}
\affiliation{ Laboratoire de Physique Statistique, Ecole Normale
Sup\'erieure, UPMC Paris 6,Universit\'e Paris Diderot, CNRS,
24 rue Lhomond, 75231 Paris Cedex 05 - France\footnote{We acknowledge the support of the ANR LHMSHE.}}

\begin{abstract}
 
We show how to apply the macroscopic fluctuation theory (MFT) of Bertini, De
Sole, Gabrielli, Jona-Lasinio, and Landim to  study the current
fluctuations of diffusive systems with a step initial condition. We argue that
one has to distinguish between two ways of averaging (the annealed and the
quenched cases) depending on whether we let  the initial condition 
fluctuate or not. Although the initial condition is not a steady state, the
distribution of the current satisfies a symmetry very reminiscent of the
fluctuation theorem. We show how the equations of the MFT can be solved in the
case of non-interacting particles. The symmetry of these equations
can be used to  deduce the distribution of the current
for  several other models, from its knowledge \cite{DG}
 for the  symmetric simple exclusion process. In the range where
the integrated current $Q_t \sim \sqrt{t}$, we show that the non-Gaussian
 decay $\exp [- Q_t^3/t]$ of the distribution of $Q_t$ is generic.
\ \\ \ \\
keywords: current fluctuations, step initial condition, fluctuation
theorem
\end{abstract}


\maketitle

\date{\today}
\ \\
{\it This work  is dedicated to our master and friend Joel Lebowitz on the occasion
of the 100th Statistical Mechanics Meeting held at Rutgers University  in December
2008}.

\section{Introduction}
\label{intro}
The study of the fluctuations of currents of energy or of particles is central
in the theory of non-equilibrium systems. Over the last decade, {\it the
macrosopic fluctuation theory} (MFT), a theory of diffusive systems maintained
in a non-equilibrium steady state by contact with two heat baths or two
reservoirs of particles, has been developed \cite{BDGJLY,BDGJLX}. This theory
was first implemented to give a framework to calculate the large deviation
functional of density profiles in non-equilibrium steady states
\cite{BDGJL1,BDGJL2,BGLeb,BGLan,KTL,Touch}. It was then understood that it
could also be used to predict the distribution of the current through
non-equilibrium diffusive systems \cite{BD,BDGJL5,BD2005,BDGJL6,BD2007,ADLW}.

The macroscopic fluctuation theory gives a large scale description of lattice
models such as the symmetric simple exclusion process (SSEP) or the Kipnis
Marchioro Presutti model \cite{KMP,HG,HG2,Imparato}. At the microcopic level,
several properties of these models can be obtained using numerical
\cite{HG,HG2,GKP}, perturbative \cite{DDR,WR}, or exact approaches
\cite{HRS1}, such as the matrix method \cite{DLS1,DLS2,ED} or the Bethe ansatz
\cite{ADLW,PM1,PM2}.
Whenever the comparison has been possible, it is remarkable that a perfect
agreement has been found between the results (on the large deviations of the
density profile \cite{DLS1,DLS2,ED} or on the probability distribution of the
current \cite{HG,HG2,DDR}) obtained by these microscopic approaches and the
predictions of the macroscopic fluctuation theory \cite{BDGJL2,BGLan,BD}.
Moreover the MFT led to the prediction of rather surprising properties of
diffusive systems, such as the possibility of phase transitions 
\cite{BD2005,BDGJL6,BD2007} in the large deviation function of the current, or
the universality of the cumulants of the current on the ring geometry
\cite{ADLW}. So far the MFT has only been used on systems at equilibrium, or
in non-equilibrium steady states.

\begin{figure}[ht]
\centerline{\includegraphics[width=8cm]{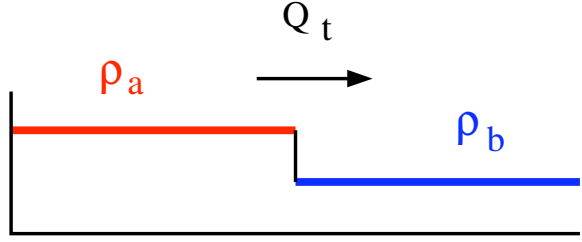}}
\caption{The step initial condition with a density $\rho_a$ at the left
of the origin  and $\rho_b$  at the right of the origin}
\label{initial}
\end{figure}

In a recent work \cite{DG}, we considered the fluctuations of the integrated
current $Q_t$ through the origin of the SSEP, starting with a non steady state
initial condition : a step in the density profile at the origin with density
$\rho_a$ on the negative axis and density $\rho_b$ on the positive axis, as
shown in figure \ref{initial}.
The SSEP is one of the simplest lattice gas models, and has been
extensively studied in the theory of non-equilibrium systems
\cite{HS,Liggett2,derrida2007}. The distribution of the integrated
current $Q_t$, for the SSEP,  is  related to the time decay of
constrained one dimensional Ising models \cite{Spohn-spin}.
In the SSEP, particles diffuse on the lattice  with
nearest neighbor jumps and a   hard core interaction which enforces that there is never  more than  one particle on each site (in practice, the configuration  at time $t$ is specified by  a binary variable $\tau_i(t)=\{1$ or $0\}$ on each lattice site, which indicates whether site $i$ is occupied or empty; the dynamics is  such that these occupation numbers are exchanged at rate 1  between every pair of neighboring sites on the lattice).
Using the Bethe  ansatz and several identities proved recently by Tracy and Widom \cite{TW1,TW2,TW3} for exclusion processes on the line,
we were able to show \cite{DG} that the generating function of the total flux $Q_t$ of particles through the origin during a long time $t$  takes the  form
 \begin{equation}
 \left\langle e^{\lambda Q_t} \right\rangle \asymp e^{\sqrt{t}  \mu(\lambda,\rho_a,\rho_b)} \,,
\label{generating}
\end{equation}
with $\mu(\lambda,\rho_a,\rho_b)$  given by
\begin{equation}
 \mu(\lambda,\rho_a,\rho_b)
 = {1 \over {\pi}} \int_{-\infty}^\infty dk \log 
\left[ 1 +\omega e^{-k^2} \right] \,,
\label{Fdef}
\end{equation}
and where $\omega$ is a function of $\rho_a, \rho_b$ and $ \lambda$
\begin{equation}
\omega= \rho_a( e^\lambda -1) + \rho_b(e^{-\lambda}- 1) + \rho_a \rho_b ( 
e^\lambda -1) ( e^{-\lambda} -1) \,.
\label{omega-def}
\end{equation}
Beyond the fact that that $\mu$ is a function of the single   parameter $\omega$, which was proved in \cite{DG}, one can see from  (\ref{generating},\ref{Fdef},\ref{omega-def}) that
\begin{enumerate}
\item All the cumulants of $Q_t$   grow like $\sqrt{t}$.
\item   $\mu(\lambda,\rho_a,\rho_b)$ satisfies a symmetry very reminiscent of the  fluctuation theorem
 \cite{ECM,GC,Kurchan,LS,Maes,Harris-S}:
\begin{equation}
  \mu \Big( \lambda,\rho_a,\rho_b \Big) =
  \mu \left(-\lambda + \log {\rho_b \over 1 - \rho_b} -  \log {\rho_a \over 1 - \rho_a} ,\rho_a,\rho_b \right) 
\label{ft}
\end{equation}
(this is because $\omega$ in (\ref{omega-def}) is left unchanged by this
symmetry).
\item For technical reasons  in the way that  (\ref{Fdef})  was derived in \cite{DG},  we had to impose the condition 
that $|\omega| <  \sqrt{2 /\pi}$. If one assumes  that  the range of validity of (\ref{Fdef})  extends to all $\omega > -1$, one gets that $ \mu \simeq {4 \over 3 \pi} (\log \omega)^{3/2}$ for large $\omega$, which would imply that   for  large  $ q $
\begin{equation}
{\rm Probability} \left( {Q_t \over  \sqrt{t}} \simeq q \right) \asymp 
\exp\left[-  {\pi^2 \over 12}  \; q^3 \; \sqrt{t}   \right] =
\exp\left[- {\pi^2 \over 12} \;  {Q_t^3 \over t} \right]\,.
\label{tail}
\end{equation}
\end{enumerate}

The goal of the present work is to see how the above results
(\ref{generating}-\ref{tail}), obtained for the SSEP with the step initial
condition of figure \ref{initial}, can be understood from the point of view of
the MFT and how they can be extended to more general diffusive systems.

\begin{figure}[ht]
\centerline{\includegraphics[width=14cm]{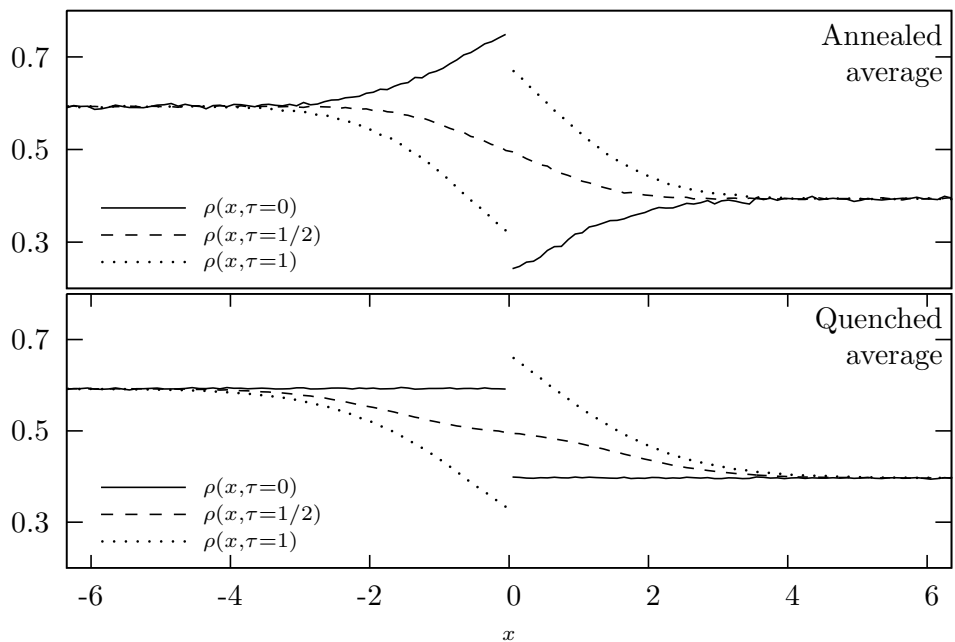}}
\caption{Average rescaled density $\rho(x,\tau)$ (see
(\ref{scaling-function-2})) for the SSEP, in the annealed and quenched cases,
when $\lambda=1.5$, $\rho_a=0.6$ and $\rho_b=0.4$. While the initial profile
is a step function in the quenched case, it deviates from it in the annealed
case. }
\label{fig:profils}
\end{figure}

When the dynamics is stochastic, the integrated current $Q_t$ through the
origin depends both on the history (i.e. on all the updates between time $0$
and time $t$) and on the initial condition (which, for the SSEP, is drawn
according to a Bernoulli measure of mean $\rho_a$ on the negative axis ($i\leq
0$) and $\rho_b$ on the positive axis ($i \geq 1$)). Very much like in the
theory of disordered systems, where one can distinguish between an {\it
annealed} average (where the partition function is averaged over all the
realizations of the disorder) and a {\it quenched} average (where the
partition function is calculated for a typical realization of the disorder),
one can define here two expressions of $\mu(\lambda)$\, :

\begin{itemize}
\item {\it the annealed case} where, as in the derivation of
(\ref{generating}-\ref{omega-def}) in \cite{DG}, one averages $e^{\lambda Q_t}$    both on the history and on the initial condition 
\begin{equation}
\mu_{\rm annnealed} (\lambda) = \lim_{t \to \infty} {1 \over \sqrt{t}}
\log  \left[ \left\langle e ^{\lambda Q_t} \right\rangle_{\rm history,  \ initial  \; condition}  \right] ;
\label{annealed}
\end{equation}
It turns out that the initial conditions which dominate the average are
atypical as shown in figure  \ref{fig:profils}.
\item {\it the quenched case}, where  one averages  $e^{\lambda Q_t}$     only on the history  for a typical   initial condition
\begin{equation}
\mu_{\rm quenched} (\lambda) = \lim_{t \to \infty}
{1 \over \sqrt{t}}
  \left\langle 
 \ \log  \left[ \left\langle e ^{\lambda Q_t} \right\rangle_{\rm history}  \right] \
\right\rangle_{\rm initial \;  condition}\,.
\label{quenched}
\end{equation}
\end{itemize}
The difference  between these two averages, and their  influence
on the distribution of the current, has already been studied for the totally asymmetric
exclusion process (TASEP) using the microscopic dynamics \cite{PS2}.

In section \ref{sec:avgs}, we formulate the calculation of both $ \mu_{ \rm
annealed}$ and $ \mu_{\rm quenched}$ in the framework of the MFT.
In section \ref{sec:symmetry}, we see that $ \mu_{ \rm annealed}$ satisfies
the symmetry (\ref{ft}) for general diffusive systems, and for general
non-steady state initial conditions. No such symmetry seems to hold in the
quenched case.
In section \ref{sec:noninter}, we consider the case of non-interacting random
walkers, where both $ \mu_{ \rm annealed}$ and $ \mu_{\rm quenched}$ can be
determined exactly.
In section \ref{sec:rotation}, we show that, for the SSEP, the
single-parameter dependence (\ref{omega-def}) of $\mu_{\rm annealed}$ can be
understood from a remarkable invariance of the MFT.
In section \ref{sec:bornes}, we obtain bounds on the decay of the distribution
of $Q_t$ which shows that (\ref{tail}) is generic for a broader class of
diffusive systems.

\section{The annealed and the quenched averages}\label{sec:avgs}
In this section we show how the macroscopic fluctuation theory
\cite{BDGJLY,BDGJLX} can be
used to calculate the generating function of the integrated current  $Q_t$ when the initial condition is a step density profile.
The theory is in principle valid for arbitrary diffusive systems with one conserved quantity,  such as the number of particles or the energy. 
 Here, for simplicity, we consider  the case of a one dimensional lattice gas  where   a configuration is  characterized by   the numbers  $n_i$ of particles on each site  $i$. 
 
Imagine first that   this one dimensional system  has a finite length $L$, and that it is in contact, at its two ends, with two reservoirs of particles at density
$\rho_a$ and $\rho_b$.
 In this finite geometry, the system's stochastic evolution
reaches a steady state,   where the flux
$Q_t$  of particles during a long $t$ has a certain average $\langle Q_t \rangle$ and a certain variance $\langle Q_t^2 \rangle_c =
\langle Q_t^2 \rangle -\langle Q_t \rangle^2$.  Close to equilibrium, i.e. when the densities of the two reservoirs  are  close  ($\rho_a \simeq \rho_b \simeq r$ with $\rho_a-\rho_b \ll r$),  and for a large system size $L$, one expects 
\cite{BD,derrida2007} that
\begin{equation}
\lim_{t \to \infty} {\langle Q_t \rangle \over t} \simeq {D(r) \over L} (\rho_a - \rho_b)
\label{D-def}
\end{equation}
and 
\begin{equation}
\lim_{t \to \infty} {\langle Q_t^2 \rangle_c \over t} \simeq {\sigma (r) \over L}\,,
\label{sigma-def}
\end{equation}
where $D(r)$ and $\sigma(r)$ are two functions which characterize the transport of particles through this diffusive system.  

At  equilibrium (for $\rho_a=\rho_b=r$), the  weights of all microscopic
configurations are  given by the Boltzmann weights.  For large $L$, if
one introduces a rescaled  position   $0<x=i/L<1$, the probability of
observing a given density profile $\rho(x)$, when the two reservoirs are
at the same density $r$, satisfies  \cite{derrida2007,BDGJLZ}  
\begin{equation}
{\rm Pro_{eq.}} (\rho(x)) \asymp \exp[- L {\cal F}_{\rm eq.} (\rho(x))  ]\,,
\nonumber
\end{equation}
where the large deviation function ${\cal F}_{\rm eq.} $ is given by  
\begin{equation}
 {\cal F}_{\rm eq.}(\rho(x))  =  \int_0^1 \Big[f(\rho(x)) - f(r) - (\rho(x) - r) f'(r) \Big] dx  \,,
\label{F-eq}
\end{equation}
and  $f(r)$ is the  free energy per site of the equilibrium system at
density $r$ (defined as 
 $f(r) = - \lim_{L \to \infty} (\log Z(L r ,L))/ L$)
 for  $Z(N,L)$ the partition function of the system with $N$ particles on $L$ sites). 
  One can show  \cite{HS,derrida2007} that   the fluctuation dissipation theorem, which is satisfied at equilibrium, implies that
 \begin{equation}
 f''(r) = {2 D(r) \over \sigma(r)}{\,.} \label{D-sigma}
\end{equation}

For a diffusive system 
on a  one dimensional lattice of $L$ sites, in contact with two reservoirs at  densities $\rho_a$ and $\rho_b$, 
the average density $\langle n_i(t) \rangle $ near position $i$
and time $t$, and
the total flux of particles $Q_i(t)$ through position $i$ between times $0$ and $t$, are expected to follow diffusive scaling laws. For large $L$, and for times of order $L^2$, they take the form
\begin{equation}\nonumber
 \langle n_i(t) \rangle  =   \rho \left( {i\over L}, {t \over L^2} \right)
\qquad {\rm and} \qquad
 Q_i(t) = L \;  q \left( {i\over L}, {t \over L^2} \right) \;. 
\end{equation}
From the large deviation hydrodynamics theory
\cite{KOV,HS,KL,BD2007,derrida2007}, the probability  of observing a
certain density profile $ \rho \left( x, \tau \right)$ and
current profile $j \left( x, \tau \right) \equiv \partial q(x,\tau) / \partial \tau$
over the rescaled time interval $0 < \tau <  t/L^2$  is expressed as
\begin{equation}
\label{eq: dev exp}
{\rm Pro} \Big( \{ \rho(x,\tau), j(x,\tau)\}  \Big)  \asymp \exp
\left[ - L \int_{0}^{t/L^2}
d \tau \int_0^1 dx  \ {\left[ j(x,\tau) + D({\rho(x,\tau}))
{\partial { \rho(x,\tau)} \over \partial x}\right]^2 \over 2
\sigma( \rho(x,\tau))} \right]{\,,}
\end{equation}
where $D(\rho)$ and $\sigma(\rho)$ are defined as in
(\ref{D-def},\ref{sigma-def}). Expression (\ref{eq: dev exp}) simply means
that, locally, Fick's law ($j= - D(\rho) \rho'$) is satisfied everywhere up to
Gaussian current fluctuations of variance $\sigma(\rho)$. The conservation of
the number of particles,\\ $ n_i(t) - n_i(0) = Q_{i-1}(t) - Q_{i}(t)$, becomes
a conservation law on the rescaled density and current profiles :
\begin{equation}
{\partial \rho \over \partial \tau} = - {\partial j \over \partial x} \,.
\label{conservation}
 \end{equation}

For a non-steady state initial condition, as in figure \ref{initial}, the
system size is infinite. If one observes the fluctuations of the current over
a long time $t$, one can introduce a characteristic length $ \sqrt{t}$. The
average density $\langle n_i(t')\rangle $ near site $i$ and the integrated
current $Q_i(t')$ between times $0$ and $t'<t$ then become scaling functions
of the form
\begin{equation}
 \langle n_i(t') \rangle  =   \rho \left( {i\over  \sqrt{t}}, {t' \over t} \right)
\qquad {\rm and} \qquad
 Q_i(t') =  \sqrt{t}  \; q \left( {i\over  \sqrt{t}}, {t' \over  t} \right) \; 
\label{scaling-function-2} \,,
\end{equation}
and the probability of observing such rescaled density  and current profiles is given by
\begin{equation}
\label{action}
{\rm Pro} \Big( \{ \rho(x,\tau), j(x,\tau)\}  \Big)  \asymp \exp
\left[ - \sqrt{t} \int_{0}^{1}
d \tau \int_{-\infty}^\infty dx  \ {\left[ j(x,\tau) + D({\rho(x,\tau}))
{\partial { \rho(x,\tau)} \over \partial x}\right]^2 \over 2
\sigma( \rho(x,\tau))} \right] \,.
\end{equation}
The integrated current $Q_t$ through the origin during time $t$    can
then be written as
\begin{equation}
Q_t = \sum_{i \geq 1} n_i(t) - n_i(0) \simeq \sqrt{t} \int_0^\infty dx  \  [\rho(x,1) - \rho(x,0) ] \,.
\label{Qt}
\end{equation}

Moreover, when the initial condition is a local equilibrium configuration at
density $\rho_a$ on the negative axis and density $\rho_b$ on the positive
axis, as in figure \ref{initial}, the probability ${\rm Pro_{\rm initial}}$ of
the initial profile $ \rho(x,0)$ is given by
\begin{equation}
\label{pro-init}
{\rm Pro_{\rm initial}} (\rho(x,0)) \asymp \exp \left[ - \sqrt{t} \; {\cal F}_{\rm init.} (\rho(x,0)) \right]\,,
\end{equation}
where (\ref{F-eq},\ref{D-sigma})
\begin{eqnarray}
\label{F-init}
 {\cal F}_{\rm init.}(\rho(x,0))  &=&  \int_{-\infty}^\infty \Big[f(\rho(x,0)) - f(r(x)) - (\rho(x,0) - r(x)) f'(r(x)) \Big] dx 
\\
 &=& \int dx \int_{r(x)}^{\rho(x,0)} dz  \ (\rho(x,0) -
z)  {2 D(z) \over \sigma(z)}\,,
\label{F-init-1}
\end{eqnarray}
and  $r(x)$ is the step density profile
\begin{equation}
r(x)=  (1 - \theta(x)) \rho_a + \theta(x) \rho_b
\label{r(x)}
\end{equation}
($\theta(x) $ is the Heaviside function).
\\ \ \\
{\bf The annealed case}

Therefore (\ref{annealed}, \ref{action},\ref{Qt}, \ref{F-init})
   lead  to the following expression for 
$\mu_{\rm annnealed}$  :
\begin{equation}
\mu_{\rm annnealed}(\lambda) = \max_{\rho(x,\tau),j(x,\tau)} \left\{
- {\cal F}_{\rm init.}(\rho(x,0))  
+ \lambda \int_0^\infty dx   \; [\rho(x,1) - \rho(x,0) ]  
-  \int_{0}^{1}
d \tau \int_{-\infty}^\infty dx \;  {\left[ j(x,\tau) + D({\rho(x,\tau}))
 {\partial\rho(x,\tau)\over\partial x}\right]^2 \over 2
\sigma( \rho(x,\tau))} \right\}
\label{mu_an}
\end{equation}
Finding the optimal  $\rho(x,\tau)$ and   $j(x,\tau)$ in (\ref{mu_an})  has to be done carefully 
because they are related by the conservation law (\ref{conservation})

As shown in Appendix  \ref{apx:hydro} (see also \cite{BDGJL2,KTL}),
 one can replace the variational form
(\ref{mu_an}) by  another  variational form :
\begin{eqnarray}
&& \mu_{\rm annnealed}(\lambda) = \max_{\rho(x,\tau),H(x,\tau)}
\left\{
- {\cal F}_{\rm init.}(\rho(x,0))
+ \lambda \int_0^\infty dx   \; [\rho(x,1) - \rho(x,0) ]
\right.  \nonumber  \\
&&  \left.
 -  \int_{0}^{1}
d \tau \int_{-\infty}^\infty dx  \left[ H(x,\tau) {\partial \rho(x,\tau)
\over \partial \tau} + D(\rho(x,\tau)) {\partial H(x,\tau)
\over \partial x}{\partial \rho(x,\tau) \over \partial x} -
{\sigma(\rho(x,\tau)) \over 2} \left( {\partial H(x,\tau) \over \partial x} 
\right)^2
\right] \right\} \,,
\label{mu_an1}
\end{eqnarray}
where $\rho(x,\tau)$ and $H(x,\tau)$ are independent.  The optimal
$\rho(x,\tau)$ and $H(x,\tau)$ in (\ref{mu_an1}) satisfy
\begin{eqnarray}\label{conservation-bis}
  {\partial \rho(x,\tau) \over \partial \tau} &=&  {\partial  \over \partial 
  x}\left[ D(\rho(x,\tau))
   {\partial  \rho(x,\tau) \over \partial x}  \right]  - {\partial  \over
  \partial x}  \left[   \sigma(\rho (x,\tau))  {\partial  H(x, \tau) \over
  \partial x}  \right]\,,\\
  \label{H-evo}
  {\partial H(x,\tau) \over \partial \tau} &=& -D(\rho(x,\tau))
    {\partial^2
    H(x, \tau) \over \partial x^2}  
  - {\sigma'(\rho(x,\tau)) \over 2} \left({\partial  H(x, \tau) \over
    \partial x} \right)^2\,,
\end{eqnarray}
with the boundary conditions 
\begin{eqnarray}  \label{t=1-bis}
  H(x,1) &=& \lambda \theta(x)\,,\\
  H(x,0) &=& \lambda \theta(x) + 2 \int_{r(x)}^{\rho(x,0)} {D(\rho) \over
  \sigma(\rho)} d \rho \,,
  \label{t=0-bis}
\end{eqnarray}
 where  we have used (\ref{D-sigma}) and (\ref{F-init}).

Thus in  annealed case one can use  either (\ref{mu_an}) or
(\ref{mu_an1}-\ref{t=0-bis}) to
obtain  $ \mu_{\rm annnealed}(\lambda) $.
Using the fact that $\rho(x,\tau)$ satisfies (\ref{conservation-bis}) and that
$\rho(x,\tau)$  and $H(x,\tau)$ have limiting values ($\rho_a,\rho_b$)
and  ($0,\lambda$) as $x\to \pm
\infty$, one can simplify  (\ref{mu_an1}) to get
\begin{equation}
\mu_{\rm annnealed}(\lambda) =
- {\cal F}_{\rm init.}(\rho(x,0))
+ \lambda \int_0^\infty dx  [\rho(x,1) - \rho(x,0) ]
-  \int_{0}^{1}
d \tau \int_{-\infty}^\infty dx  {\sigma(\rho(x,\tau)) \over 2}
\left({\partial { H(x,\tau)} \over \partial x}\right)^2
\label{mu_an-bis}
\end{equation}

\ \\ \ \\
{\bf The quenched case}

In the quenched case, the main difference is that $\rho(x,0)$ is no longer allowed to fluctuate. Therefore 
 the boundary condition 
  (\ref{t=0-bis})
at $\tau=0$   is replaced by  
\begin{equation} 
\rho(x,0) = r(x)\,,
\label{t=0-ter}
\end{equation}
and (\ref{mu_an}) becomes
\begin{equation}
\mu_{\rm quenched}(\lambda) = \max_{\rho(x,\tau),j(x,\tau)} \left\{
 \lambda \int_0^\infty dx   \; [\rho(x,1) - \rho(x,0) ]  
-  \int_{0}^{1}
d \tau \int_{-\infty}^\infty dx \;  {\left[ j(x,\tau) + D(\rho(x,\tau))
 {\partial\rho(x,\tau)\over\partial x}\right]^2 \over 2
\sigma( \rho(x,\tau))} \right\}{\,,}
\label{mu_qu}
\end{equation}
with the $\max$ taken over all the profiles satisfying (\ref{conservation})
and (\ref{t=0-ter}). In terms of the field $H$, one gets
\begin{equation}
\mu_{\rm quenched}(\lambda) = 
 \lambda \int_0^\infty dx  [\rho(x,1) - \rho(x,0) ]  
-  \int_{0}^{1}
d \tau \int_{-\infty}^\infty dx  {\sigma(\rho(x,\tau) \over 2} 
\left({\partial { H(x,\tau)} \over \partial x}\right)^2 
\label{mu_qu-bis}\,,
\end{equation}
 with $\rho$ and $H$ satisfying (\ref{conservation-bis}-\ref{t=1-bis}) but with
(\ref{t=0-ter} ) instead of   (\ref{t=0-bis}).
\ \\ \ \\ \ \\ \ \\
\ 
{\bf Remark :}
From the expressions (\ref{mu_an},\ref{mu_qu}) and 
(\ref{F-init-1}),
$${\cal F}_{\rm init}(\rho(x,0)) = \int dx \int_{r(x)}^{\rho(x,0)} dz  \ (\rho(x,0) -
z)  {2 D(z) \over \sigma(z)}\,,$$
one can see that  the case  where $D(\rho)=1$ and $\sigma(\rho)$ is  a
quadratic function \cite{BGLeb,Imparato} of $\rho$, 
\begin{equation}
D(\rho)=1 \ \ \ \ ; \ \ \ \  \sigma(\rho) = 2 A \rho (B- \rho)\,,
\label{D-sigma-quadratic}
\end{equation}
 can be easily related to the SSEP, for which (see \cite{BD})
\begin{equation}
D(\rho)=1 \ \ \ \ ; \ \ \ \  \sigma(\rho) = 2  \rho (1- \rho)\,.
\label{D-sigma-SSEP}
\end{equation}
In fact, if one makes the change of variable
$$ \rho \to B \rho \ \  \ \  ;  \ \ \ \ j \to B j $$
one gets for the choice (\ref{D-sigma-quadratic}) that,  both in the
annealed and in the quenched case,
\begin{equation}\label{mu-quadratic}
\mu(\lambda,\rho_a,\rho_b) = {1 \over A}  \mu_{SSEP} \left( A B \lambda,
{\rho_a \over B},{\rho_b \over B} \right)\,.
\end{equation}
In the annealed case, where   the exact expression   of the SSEP is available
(\ref{generating},\ref{Fdef},\ref{omega-def}), one gets 
\begin{equation}
\mu_{\rm annealed} (\lambda,\rho_a,\rho_b) 
 = {1 \over  A {\pi}} \int_{-\infty}^\infty dk \log
\left[ 1 +\omega e^{-k^2} \right] \,,
\label{mu-an-quadratic}
\end{equation}
where 
\begin{equation}
\omega= {\rho_a( e^{A B \lambda} -1) \over B} +{ \rho_b(e^{- A B \lambda
}- 1) \over B} + {\rho_a \rho_b (
e^{A B \lambda} -1) ( e^{- A B \lambda} -1) \over B^2} \,.
\label{omega-def-1}
\end{equation}
 In the limit  $ A = B^{-1} \to 0$, 
this gives $\mu_{\rm annealed}$ 
when $\sigma= 2 \rho$, i.e. in 
the case of non-interacting particles (\ref{mu_an-ter3})  
that we will discuss in  section \ref{sec:noninter}.
Assuming that (\ref{mu-an-quadratic},\ref{omega-def-1}) remain valid for
non-physical values of $A$ and $B$, one would get $\mu_{\rm annealed}$,
without any further calculation, for the Kipnis Marchioro Presutti model
\cite{KMP,HG,HG2,Imparato} where $\sigma=4 \rho^2$ (in the limit $B \to 0,
A\to -2$).

\section{The time reversal symmetry}\label{sec:symmetry}
In this section we are going to see that the symmetry (\ref{ft}) can be extended to  more general diffusive systems.
To do so, let us  consider the difference  $ {\cal F}_{\rm init.}(\rho(x,1)) - {\cal F}_{\rm init.}(\rho(x,0))$. Using (\ref{F-init}), one has
\begin{equation}
\nonumber
 {\cal F}_{\rm init.}(\rho(x,1))
 -{\cal F}_{\rm init.}(\rho(x,0))
  =  \int_{-\infty}^\infty \Big[f(\rho(x,1)) - f(\rho(x,0)) - (\rho(x,1) - \rho(x,0)) f'(r(x)) \Big] dx \,,
\end{equation}
which can be rewritten as
\begin{equation}
\nonumber       
 {\cal F}_{\rm init.}(\rho(x,1))
 -{\cal F}_{\rm init.}(\rho(x,0))
  = 
 \int_0^1 d \tau
 \int_{-\infty}^\infty dx 
 f'(\rho(x, \tau))   {\partial \rho(x,\tau)  \over d \tau}  
- \int_{-\infty}^\infty   (\rho(x,1) - \rho(x,0)) f'(r(x))  dx  \,.
\end{equation}
Then,  using (\ref{conservation}), an integration by parts,  and (\ref{D-sigma}), one gets
\begin{eqnarray*}
 {\cal F}_{\rm init.}(\rho(x,1))
 -{\cal F}_{\rm init.}(\rho(x,0))
  =  &&
 \int_0^1 d \tau \int_{-\infty}^\infty dx 
f''(\rho(x, \tau)) {\partial \rho(x, \tau) 
\over \partial x}   
 j(x,\tau)
- \int_{-\infty}^\infty   (\rho(x,1) - \rho(x,0)) f'(r(x))  dx  
 \\  = && 
 \int_0^1 d \tau \int_{-\infty}^\infty dx 
{2 D(\rho(x, \tau)) \over \sigma(\rho(x, \tau))} {\partial \rho(x, \tau) 
\over \partial x}    j(x,\tau)
- \int_{-\infty}^\infty   (\rho(x,1) - \rho(x,0)) f'(r(x))  dx  \,.
\end{eqnarray*}
This allows one to rewrite the last term in (\ref{mu_an})  as
\begin{equation}
\iint
d \tau dx \;  {\left[ j + D
 \partial_x\rho\right]^2 \over 2
\sigma} = 
  { {\cal F}_{\rm init.}(\rho(x,1))
 -{\cal F}_{\rm init.}(\rho(x,0)) \over 2}
+ \int_{-\infty}^\infty   (\rho(x,1) - \rho(x,0)) {f'(r(x))   \over 2} dx  
+\iint d \tau  dx \;  { j^2 + (D\partial_x\rho)^2 \over 2 \sigma}  
\label{identity}
\end{equation}
and therefore  (\ref{mu_an}) becomes
\begin{eqnarray}
\mu_{\rm annnealed}(\lambda) = \max_{\rho(x,\tau),j(x,\tau)} \left\{
  -{ {\cal F}_{\rm init.}(\rho(x,1))
 +{\cal F}_{\rm init.}(\rho(x,0)) \over 2}
+  \int_{-\infty} ^\infty dx   \; [\rho(x,1) - \rho(x,0) ]  \left[\lambda \theta(x) - {f'(r(x)) \over 2} \right]
 \right.  \nonumber  \\  \left. -  
\int_{0}^{1}
d \tau \int_{-\infty}^\infty dx \;  { j(x,\tau)^2 + \left(D(\rho(x,\tau))
 \partial_x\rho(x,\tau)\right)^2 \over 2
\sigma( \rho(x,\tau))} \right\}\,.  &&
\label{ft-4}
\end{eqnarray}
For the step initial density profile (\ref{r(x)}), one has (\ref{D-sigma}) 
$$f'(r(x)) = f'(\rho_a) -  \theta(x)  \int_{\rho_b}^{\rho_a} { 2 D(\rho) \over  \sigma(\rho)} d \rho\,.$$
One can then see in (\ref{ft-4})  that the initial time and the final time play  symmetric roles : if one replaces  $\{\rho(x,\tau),j(x,\tau) \}$  by
$\{\rho(x,1-\tau),-j(x,1-\tau) \}$, (\ref{ft-4}) is left unchanged provided that $ \lambda \to - \lambda -  \int_{\rho_b}^{\rho_a} { 2 D(\rho) \over  \sigma(\rho)} d \rho$.
\Big(one has to use that,  from the conservation of the total number of particles, $
\int_{-\infty}^\infty [ \rho(x,1) - \rho(x,0)] dx = 0 $\Big).
Therefore $\mu_{\rm annealed}$ satisfies
\begin{equation}
\mu_{\rm annealed} \big( \lambda \big) = 
\mu_{\rm annealed} \Big(  - \lambda -    \int_{\rho_b}^{\rho_a} { 2 D(\rho) \over  \sigma(\rho)} d \rho \Big) \,.
\label{ft-5}
\end{equation}
This is a generalization of (\ref{ft}) (for the SSEP (\ref{D-sigma-SSEP}), $D(\rho)=1$ and
$\sigma(\rho)=2 \rho(1-\rho)$, and (\ref{ft-5})
reduces to (\ref{ft})) and therefore shows that a version \cite{derrida2007}
of the fluctuation theorem \cite{ECM,GC,Kurchan,LS,Maes,Harris-S} holds for
general diffusive systems with the step initial condition considered here.
Although this initial condition is neither an equilibrium state, nor a
non-equilibrium steady state, the time reversal symmetry (\ref{ft-5}) holds.
We think that this is because, in the annealed case, the initial condition is
in local equilibrium.

One can repeat  the same   transformations in the quenched case.  Due to the  absence   of ${\cal F}_{\rm init.}(\rho(x,0)) $  in  (\ref{mu_qu}),  one  ends up with an expression  where $\rho(x,1)$ and $\rho(x,0)$  do not play symmetric roles, 
so  that $\mu_{\rm quenched}$ does not seem to satisfy any kind of time reversal symmetry.
\ \\ \ \\
 {\bf Remark :} By the same reasoning, one can show that the symmetry  (\ref{ft-5}) holds for initial conditions more general than the step initial profile.
One can consider at $t=0$ an initial density profile
\begin{equation}
r(x)= (1-v(x))\rho_a + v(x) \rho_b \,,
\label{r(x)-bis}
\end{equation}
where $v(x)$ is no longer a step as in (\ref{r(x)}) but could be a more
general sigmoid function with $v(-\infty) = 0$ and $v(\infty)=1$. One can also
replace the measure of the integrated current (\ref{Qt}) at the origin by its
weighted average over space in a region around the origin :
\begin{equation}
Q_t = \sqrt{t} \int_{-\infty}^\infty dx  \; w(x) [\rho(x,1) - \rho(x,0) ] \,, 
\nonumber
\end{equation} where $w(x)$ is another sigmoid function. Then following exactly the same steps as in the derivation of  (\ref{ft-4})
one gets 
\begin{eqnarray}
\mu_{\rm annnealed}(\lambda) = \max_{\rho(x,\tau),j(x,\tau)} \left\{
  - { {\cal F}_{\rm init.}(\rho(x,1))
 +{\cal F}_{\rm init.}(\rho(x,0)) \over 2}
+  \int_{-\infty} ^\infty dx   \; [\rho(x,1) - \rho(x,0) ]  \left[\lambda w(x) + \int_{r(x)}^{\rho_a} {D(\rho) \over \sigma(\rho)} d \rho \right]
 \right.  \nonumber  \\  \left. -  
\int_{0}^{1}
d \tau \int_{-\infty}^\infty dx \;  { j(x,\tau)^2 + (D(\rho(x,\tau))
 \partial_x\rho(x,\tau))^2 \over 2
\sigma( \rho(x,\tau))} \right\}\,,\;\,  &&
\nonumber
\end{eqnarray}
from which one can see that the time reversal symmetry  (\ref{ft-5}) remains valid if $v(x)$ and $w(x)$ are related by
\begin{equation}\label{cond-symmetry}
  w(x) = \left[ \int_{r(x)}^{\rho_a} {D(\rho) \over \sigma(\rho)} d \rho
  \right] / \left[\int_{\rho_b}^{\rho_a} {D(\rho) \over \sigma(\rho)} d \rho
  \right] \,.
\end{equation}

{\bf Remark :}  No time reversal symmetry seems to hold in the quenched case.
However, if an additional symmetry (the particle-hole symmetry) holds,
one can relate $\mu_{\rm annealed}$ and $\mu_{\rm quenched}$.
In Appendix \ref{apx:sym}, we show
that,
if $D(\rho)$ and $\sigma(\rho)$ satisfy
\begin{equation}\label{D-sigma-sym}
  \left\lbrace \begin{array}{lll}
    D(\rho)& = & D(2\bar\rho-\rho) \\
    \sigma(\rho) &=& \sigma(2\bar\rho-\rho)
  \end{array}\right.\,,
\end{equation}
then the optimal profile $\rho^{\rm (a)}(x,\tau)$ for the annealed
variational problem (\ref{mu_an}) when  $\rho_a=\rho_b=\bar\rho$ is such
that
\begin{equation}
\rho^{\rm (a)}(x,\tau) =  2 \bar \rho - \rho^{\rm (a)}(x,1-\tau)  \ .
\label{part-hole-sym}
\end{equation}  This implies in particular that  $\rho^{\rm (a)}(x,\tau=1/2)= \bar\rho$ and    allows  one to relate  the optimal  annealed (\ref{mu_an})    and quenched  (\ref{mu_qu}) profiles 
(see  Appendix \ref{apx:sym}),  leading to 
\begin{equation}
\mu_{\rm quenched}(\lambda,\rho_a=\rho_b=\bar\rho)={1\over\sqrt{2}}
\mu_{\rm annealed}(\lambda,\rho_a=\rho_b=\bar\rho)\,.
\label{part-hole-sym1}
\end{equation}
For the  SSEP (\ref{D-sigma-SSEP}) the particle-hole  symmetry (\ref{D-sigma-sym}) is satisfied, and therefore (\ref{part-hole-sym1}) holds, for $\bar\rho=1/2$.
 Thus\\ $\mu_{\rm quenched}(\lambda,\rho_a=\rho_b
=1/2)$ can be deduced from the exact expression
(\ref{Fdef},\ref{omega-def}).

\section{The non interacting walkers}\label{sec:noninter}
The problem with the expressions (\ref{mu_an}) or (\ref{mu_qu}) is that it is
very hard to solve the equations satisfied by the time dependent density and
current profiles for general $D(\rho)$ and $\sigma(\rho)$. In this section, we
solve the easy case of non-interacting random walkers.

Let us consider  non-interacting particles on an infinite one dimensional
lattice.   Each   paticle on this lattice  jumps  at rate   1   to  each
of its neighboring sites, irrespective of the positions of the other
particles. 
One can show (see appendix \ref{gaussien}) that in this case
\begin{equation}
D(\rho)=1 \ \ \ ; \ \ \  \sigma(\rho)=2 \rho \ \ \ ; \ \ \ f(\rho)= \rho
\log \rho - \rho \,.
\label{D-sigma-f-gaussien}
\end{equation}
Then  (\ref{conservation-bis})  becomes
\begin{equation}
{\partial \rho(x,\tau) \over \partial \tau} =  
 {\partial^2  \rho(x,\tau) \over \partial x^2}    -  {\partial  \over
\partial x}  \left[  2  \rho (x,\tau)  {\partial  H(x, \tau) \over
\partial x}  \right] \,,
\label{conservation-bis-gaussien}
 \end{equation}
and the evolution equation (\ref{H-evo})  of $H$ becomes
autonomous :
\begin{equation}
{\partial H(x,\tau) \over \partial \tau} = -
  {\partial^2
  H(x, \tau) \over \partial x^2}
-  \left({\partial  H(x, \tau) \over
  \partial x} \right)^2 \,.
\label{H-evo-gaussien}
 \end{equation}
It is easy to check that the  general solution of
(\ref{conservation-bis-gaussien},\ref{H-evo-gaussien}) can be written as
\begin{equation}
 H(x,\tau)=  K + \ln G(x,\tau) \ \ \ \ \ {\rm and} \ \ \ \ \  \rho(x,\tau)= G(x,\tau) R(x,\tau) \,,
\nonumber
\end{equation}
where $G$ antidiffuses and $R$ diffuses :
\begin{equation}
 {\partial G(x,\tau) \over \partial \tau} =  
- {\partial^2 G(x,\tau) \over \partial x^2}  
\ \  \ \ \ ; \ \ \ \ \  
 {\partial R(x,\tau) \over \partial \tau} =  
 {\partial^2 R(x,\tau) \over \partial x^2}\,.
\nonumber
\end{equation}
As the boundary condition
(\ref{t=1-bis})
 holds both for the annealed and the quenched case, one gets
$$G(x,\tau)= {e^{\lambda} + 1 \over 2} + {e^{\lambda} - 1  \over 2}
E\left(x \over 2 \sqrt{1-\tau} \right)\,,  $$
where $E(z)$ is the error function
\begin{equation}
\label{error}
E(z)= {2 \over \sqrt{\pi}} \int_0^z e^{-u^2} du \,.
\end{equation}
\
{\bf In the annealed case}, the boundary condition (\ref{t=0-bis}) becomes
$$\rho(x,0) =  r(x) e^{- \lambda \theta(x)} G(x,0)$$
and, 
 using (\ref{D-sigma-f-gaussien}),
the solution of (\ref{conservation-bis-gaussien})
 for this
boundary condition is 
\begin{equation}
\rho(x,\tau)= \left[ {\rho_b e^{-\lambda}  + \rho_a\over 2 }+
{ \rho_b e^{-\lambda}  - \rho_a \over 2 } E\left(x \over 2 \sqrt{\tau} \right) \right]
G(x,\tau) \,.
\nonumber
\end{equation}

Using (\ref{conservation-bis},\ref{H-evo},\ref{D-sigma-f-gaussien}), one can show that
\begin{equation}
\rho \left( {\partial H \over \partial x} \right)^2
={\partial (H \rho) \over \partial \tau} - 
  {\partial \over \partial x} \left(  H {\partial \rho \over \partial x} - \rho  {\partial H \over \partial x}  - 2 H \rho  {\partial H \over \partial x} \right)  \,.
\nonumber
\end{equation}
Using this identity in (\ref{mu_an-bis}) and the fact that $\rho$ and $H$ have limiting values at $\pm \infty$, one gets 
\begin{equation}
\mu_{\rm annnealed}(\lambda) =
- {\cal F}_{\rm init.}(\rho(x,0))
+ \lambda \int_0^\infty dx  [\rho(x,1) - \rho(x,0) ]
-  
 \int_{-\infty}^\infty dx  [H(x,1) \rho(x,1) - H(x,0) \rho(x,0)] \,.
\label{mu_an-ter}
\end{equation}
Using (\ref{D-sigma},\ref{F-init},\ref{t=1-bis}) and (\ref{t=0-bis}), one then 
has
\begin{equation}
\mu_{\rm annnealed}(\lambda) =
 \int_{-\infty}^\infty dx   [\rho(x,0) f'(\rho(x,0)) - f(\rho(x,0))
-r(x) f'(r(x)) + f(r(x))]\,,
\nonumber
\end{equation}
and, as $f(\rho) = \rho \log \rho - \rho$, one gets
\begin{equation}
\mu_{\rm annnealed}(\lambda) =
 \int_{-\infty}^\infty dx   [\rho(x,0)  - r(x)]
 =
 \rho_a {e^\lambda -1 \over 2} \int_{-\infty}^0 \left[ 1 + E \left( {x \over 2} \right) \right] dx
 + \rho_b {e^{-\lambda} -1 \over 2} \int_0^{\infty} \left[ 1 - E \left( {x \over 2} \right) \right] dx \,.
\nonumber
\end{equation}
 This leads to
\begin{equation}
\mu_{\rm annnealed}(\lambda) =
{ \rho_a (e^\lambda -1) +  \rho_b (e^{-\lambda} -1) \over  \sqrt{\pi}}\,. 
\label{mu_an-ter3}
\end{equation}
One can notice that this is just the limit of
(\ref{Fdef},\ref{omega-def}) when $\rho_a$ and $\rho_b$ are small (at low
density the exclusion rule in the SSEP can be neglected).
One can also see  by expanding (\ref{mu_an-ter}) in powers of $\lambda$ that
in the long time limit
\begin{equation}
\label{Q-annealed}
{\langle Q \rangle \over t} \to {\rho_a - \rho_b \over \sqrt{\pi}}
\ \ \ \ \ ; \ \ \ \ 
{\langle Q^2 \rangle_c \over t} \to {\rho_a + \rho_b \over \sqrt{\pi}}\,.
\end{equation}
\\ \ \\ \ \\ \
\
{\bf In the quenched case}, the boundary condition is (\ref{t=0-ter}) instead of 
(\ref{t=0-bis}). Therefore the
profile becomes
\begin{equation}
\rho(x,\tau)= \left[ {\rho_b   + \rho_a\over 2 }+
{ \rho_b   - \rho_a \over 2 } E\left(x \over 2 \sqrt{\tau} \right) \right]
{G(x,\tau) \over G(x,0)} \,.
\nonumber
\end{equation}
Then, following the same steps as in the derivation of  (\ref{mu_an-ter}), one gets
\begin{equation}
\mu_{\rm quenched}(\lambda) =
 \lambda \int_0^\infty dx  [\rho(x,1) - \rho(x,0) ]
-  
 \int_{-\infty}^\infty dx  [H(x,1) \rho(x,1) - H(x,0) \rho(x,0)]\,,
\nonumber
\end{equation}
which leads to 
\begin{equation}
\mu_{\rm quenched}(\lambda) =
\rho_a \int_{-\infty}^0  dx \log  G(x,0) \ \ + \ \ \rho_b \int_0^{\infty}  dx \log[ e^{-\lambda}   G(x,0) ]\,.
\nonumber
\end{equation}
Therefore
\begin{equation}
\mu_{\rm quenched}(\lambda) =
\rho_a \int_{-\infty}^0  dx 
\log  \left[ {e^{\lambda} + 1 \over 2} + {e^{\lambda} - 1  \over 2}
E\left(x \over 2 \right) \right]
 \ \ + \ \ \rho_b \int_0^{\infty}  dx 
\log  \left[ {1+ e^{-\lambda}  \over 2} + { 1- e^{-\lambda}   \over 2}
E\left(x \over 2  \right) \right]\,.
\label{mu_qu-ter2}
\end{equation}
The expansion in powers of $\lambda$ leads to 
\begin{equation}
\label{Q-quenched}
{\langle Q \rangle \over t} \to {\rho_a - \rho_b \over \sqrt{\pi}}
\ \ \ \ \ ; \ \ \ \ 
{\langle Q^2 \rangle_c \over t} \to {\rho_a + \rho_b \over \sqrt{2 \pi}}\,,
\end{equation}
which shows that the annealed (\ref{Q-annealed}) and quenched
(\ref{Q-quenched}) cases start to differ at the level of the variance of
$Q_t$.
\ \\ \ \\ \ \\ \ \\
{\bf Remark :}
Taking the $\lambda \to \infty$ limit of the expression  (\ref{mu_qu-ter2}) of $\mu_{\rm quenched}(\lambda)$, one obtains
$$\mu_{\rm quenched} (\lambda) \underset{\lambda\to\infty}{\sim}
{4 \over 3}\rho_a\lambda^{3/2}\,,$$
(with a similar result with $\rho_b$ replaced by  $\rho_a$ and $\lambda$ by $|\lambda|$ for $\lambda\to
-\infty$). Then, we can perform a Legendre transform to obtain the
decay of the distribution of the integrated current $Q_t$, as defined in
(\ref{quenched}), which yields
\begin{equation}\label{tail-gauss}
  {\rm Pro}\left[{Q_t\over\sqrt{t}}\simeq q\right] \underset{q\to\infty}{\asymp}
  \exp\left[-{\sqrt{t}q^3\over 12\rho_a^2}\right]\,.
\end{equation}
This non-Gaussian decay is very reminiscent of 
the SSEP (\ref{tail}). In section \ref{sec:bornes}, we will show that
 this type of  decay is rather generic.

Expression (\ref{tail-gauss}) can alternatively be understood  from (\ref{Pij}), as the  tail is dominated by the contribution of  the first $Q_t$ particles at the left of the origin, that is :
$${\rm Pro}\left[{Q_t\over\sqrt{t}}\simeq q\right]  \asymp \prod_{i=1}^{Q_t} \exp \left[  - {i^2 \over  4 t \rho_a^2} \right]  \asymp \exp \left[- {Q_t^3 \over 12 t \rho_a^2} \right]\,,
$$
where we have used that the average distance between consecutive particles is $1/\rho_a$. In the annealed case where the initial profile can fluctuate, the decay is slower, because the events which dominate have an initial profile where the $Q_t$ particles are arbitrarily close to the origin.

\def\sh{\mathrm{\,sinh\, }}
\def\ch{\mathrm{\,cosh\, }}
\section{Rotational symmetry for the SSEP}
\label{sec:rotation}
In this section, we consider the MFT of the symmetric simple
exclusion process (SSEP), for which (\ref{D-sigma-SSEP}) $D(\rho)=1$ and
$\sigma(\rho)=2\rho(1-\rho)$.
The MFT then exhibits a remarkable symmetry : in the annealed case, this
symmetry allows us to relate the generating functions of the integrated
current $Q_t$ for different values of the initial densities $\rho_a$ and
$\rho_b$. This relationship takes the form of the single-parameter dependence
(\ref{Fdef})
$$\mu_{\rm annealed}(\lambda,\rho_a,\rho_b) =
F(\omega(\lambda,\rho_a,\rho_b))\,,$$
with $\omega$  given by
(\ref{omega-def}). 
 This $\omega$ dependence was already derived
by considering the microscopic dynamics of the SSEP in \cite{DG}. Here, it is
recovered by showing that, when $\omega(\lambda,\rho_a,\rho_b) =
\omega(\lambda',\rho'_a,\rho'_b)$, an explicit transform relates the
variational problems (\ref{mu_an}) with parameters $(\lambda,\rho_a,\rho_b)$
and $(\lambda',\rho'_a,\rho'_b)$. 

This transform is inspired by a known representation of the microscopic
exclusion process \cite{KTL} in terms of spins : here, the equivalent of
a global rotation of these spins will allow us to go from $(\lambda,\rho_a,
\rho_b)$ to $(\lambda',\rho'_a,\rho'_b)$. When $\mu$ is
expressed as an optimum (\ref{mu_an1}) over the two independent variables
$\rho(x,t)$ and $H(x,t)$, one can introduce a "spin" variable,

\begin{equation}\nonumber
  \left\lbrace \begin{array}{ll}
    S_+ &= \rho\, e^{-H}\\
    S_- &= (1-\rho) e^{H}\\
    S_z &= \rho-{1\over 2}
  \end{array}\right. \mbox{ , and the quadratic form } \vec S \cdot \vec S' = 
  {1\over2}(S_+S'_- + S_- S'_+)+S_z^2\,.
\end{equation}

The bulk term in the variational problem (\ref{mu_an1}) can then be rewritten as

\begin{equation}
\iint -H\partial_\tau\rho - \partial_x H\partial_x \rho +\rho(1-\rho)(\partial_xH)^2 = \iint -H\partial_\tau\rho
-\partial_x\vec S \cdot \partial_x\vec S\,. 
\label{spindef}
\end{equation}

The last term of this "action", $-\partial_x\vec S \cdot
\partial_x\vec S$, is clearly invariant under orthogonal transforms of $\vec S$.
Thus, starting from the optimal profiles $(\rho,H)$ for a given set of
parameters $(\lambda,\rho_a,\rho_b)$, one can deduce sets of profiles
$(\rho',H')$, obtained by performing an orthogonal transform on $\vec S$, 
which satisfy the same bulk minimization equations (\ref{conservation-bis},
\ref{H-evo}) as $(\rho,H)$.

Therefore, for $(\rho',H')$ to be the optimal profiles for other values of the
parameters $(\lambda',\rho'_a,\rho'_b)$, it is sufficient that they satisfy
the corresponding boundary conditions : (\ref{t=1-bis},\ref{t=0-bis}) in the
annealed case, and (\ref{t=1-bis},\ref{t=0-ter}) in the quenched case. 

Let us first look at these boundary conditions at $\tau=0,1$ for $x\to
\pm\infty$ :
$$\left\lbrace \begin{array}{l}
  \rho(-\infty,\tau)=\rho_a\\
  H(-\infty,\tau)=0\\
\end{array}\right.\,\,\, \mbox{ as well as }\,\,\left\lbrace \begin{array}{l}
    \rho(\infty,\tau)=\rho_b\\
    H(\infty,\tau)=\lambda\\
  \end{array}\right.\,,$$
which correspond to $\vec S_{-\infty}=(\rho_a,1-\rho_a,\rho_a-1/2)$ and
$\vec S_{+\infty}=(\rho_b e^{-\lambda},(1-\rho_b)e^{\lambda},\rho_b-1/2)$.
Under an orthogonal transform on $\vec S$, the scalar product of these vectors
is necessarily conserved :
$$\vec S_{-\infty} \cdot \vec S_{+\infty} ={1\over2}\left(
\rho_a\left(1-\rho_b\right)e^{\lambda}+\rho_b
e^{-\lambda}\left(1-\rho_a\right)\right)+\left(\rho_a-{1\over 2}\right)\left(\rho_b-{1\over 2}\right) = {\omega\over2}+{1\over4}\,,$$
with $\omega$ as defined in (\ref{omega-def}). Hence $\omega=\omega'$
is a necessary condition for $(\rho',H')$ to be optimal for the set of
parameters $(\lambda',\rho'_a,\rho'_b)$.

In order to explicitly check that one can indeed relate the optimal profiles
when $\omega=\omega'$, and to compare the corresponding generating functions,
we will now express the optimal profiles $(\rho,H)$ for
$(\lambda,\rho_a,\rho_b)$ in terms of the "reference profiles" 
$(\tilde\rho,\tilde H)$ obtained for the SSEP at uniform density $1/2$ : $\tilde\rho_a=\tilde\rho_b=1/2$. When 
 $\omega=\tilde\omega$, we reparametrize $\rho_a$ and $\rho_b$ in terms of two variables $u$ and $v$ :
$$ \rho_a ={e^v\ch u-1\over e^\lambda-1} \,\,\,\,\mbox{ and }\,\,\,\,
  \rho_b={e^{-v}\ch u-1\over e^{-\lambda}-1}\,,$$
so that $$\omega=\sh^2 u  \ \ \ \  {\rm and} \ \ \ \  \tilde\lambda=2 u \, .$$
 One can then check (after some algebra) that
the mapping $(\tilde\rho,\tilde H)\to (\rho,H)$, as
\begin{equation}\label{rotation}
  \left\lbrace \begin{array}{ll}
  \rho&={1\over \sh u \sh {\lambda\over 2}}\left(e^{\tilde H-u}\sh{\lambda+u-v\over2}-\sh {\lambda-u-v\over2}
  \right)\left(\tilde\rho e^{u-\tilde H}\sh{u+v\over 2}-(1-\tilde\rho) 
  \sh{u-v\over2} \right)\\

    e^{H} &= 1 + {e^u(e^{\lambda}-1)(e^{\tilde H}-1)\over
    e^{\tilde H}(e^u-e^v)+e^u(e^{u+v}-1)}\\
  \end{array}\right.\,,
\end{equation}
gives a solution of the bulk equations (\ref{conservation-bis},\ref{H-evo}).

From the expression of $e^{H}$, one can easily see that the final time
boundary condition (\ref{t=1-bis}), which is common to the annealed 
and quenched  cases, carries over from $\tilde H$
to $H$ :
\begin{equation}\label{CFSSEP}
   \tilde H(x,1)=2u\,\theta(x)\;\Longrightarrow\;
  H(x,1)=\lambda\theta(x)\,.
\end{equation}

However, the initial-time boundary condition behaves differently in the
annealed and in the quenched cases. In the quenched case, one would need that
$\rho(x,0)=r(x)$ when $\tilde \rho(x,0)=1/2$ : this requires (\ref{rotation})
that $\tilde H(x,0)=\tilde\lambda\theta(x)$, which is not expected to be
satisfied as $\tilde H(x,0)$ is free under the quenched boundary conditions.
Hence the condition (\ref{t=0-ter}) does not carry over from
$(\tilde\rho,\tilde H)$ to $(\rho,H)$, and (\ref{rotation}) does not lead to
the correct optimal profiles in the quenched case.

On the other hand, the initial-time condition in the annealed case
(\ref{t=0-bis}) is $\tilde H(x,0)=2u\theta(x) +f'(\tilde \rho(x,0)) -
f'(1/2)$.
Integrating the Einstein relationship (\ref{D-sigma}) for $D=1$,
$\sigma=2\rho(1-\rho)$ leads to
\begin{equation}\label{fsigquad}
  f'(r) = \log{r\over 1-r} \;\;\mbox{and}\;\;
  f(r)=r\log r+(1-r)\log(1-r)\,.
\end{equation}
One can then check that (\ref{rotation}) yields
\begin{equation}\label{CISSEP}
\tilde H(x,0)=2u\,\theta(x)+\log{\tilde \rho(x,0)\over1-\tilde \rho(x,0)}\;
\Longrightarrow\;H(x,0)=\lambda\theta(x) + \log { \rho(x,0) \over 1-  \rho(x,0)}-\log{r(x)\over 1-r(x)}\,.
\end{equation}
Therefore (\ref{rotation}) maps the optimal profiles for
$(\tilde\lambda,1/2,1/2)$ to those for $(\lambda,\rho_a,\rho_b)$ in the
annealed case.

 This in turn allows us to relate the generating functionals
$\mu_{\rm annealed}(\tilde\lambda,1/2,1/2)$ and $\mu_{\rm
annealed}(\lambda,\rho_a,\rho_b)$ : taking into account the invariance of
the bulk term, we obtain from (\ref{mu_an1},\ref{spindef})
\begin{eqnarray*}
  \mu_{\rm an.}(\lambda,\rho_a,\rho_b)-\mu_{\rm an.}(\tilde\lambda,1/2,1/2) &=&\int_0^\infty dx\left[\lambda(\rho(x,1) -\rho(x,0))-\tilde{\lambda}(\tilde\rho(x,1)-\tilde\rho(x,0))\right]\\
  & &-{\cal F}_{init.}(\rho(x,0))+\tilde{{\cal F}}_{init.}(\tilde\rho(x,0))\\
  & &-\iint d\tau dx [H\partial_\tau \rho - \tilde H\partial_\tau
  \tilde\rho]\,.
\end{eqnarray*}
Integrating by parts the last term and using (\ref{CFSSEP}-\ref{CISSEP}), this can be simplified to
\begin{equation}\label{act-diff}
  \mu_{\rm an.}(\lambda,\rho_a,\rho_b)-\mu_{\rm an.}(\tilde\lambda,1/2,1/2)=
  \int dx\log{1-r(x)\over 1-\rho(x,0)}{1-\tilde\rho(x,0)\over 1-1/2} + \iint
  d\tau dx [\rho\partial_\tau H-\tilde\rho\partial_\tau\tilde H]\,.
\end{equation}
From (\ref{rotation}), one can express $ \rho \partial_\tau H -
\tilde\rho \partial_\tau \tilde H $ as a total derivative in terms of $ \tilde
H$ :
$$\rho \partial_\tau H -\tilde\rho \partial_\tau \tilde H = - {
\partial \over \partial \tau } \log \left[ ( e^u - e^v ) e^{ \tilde H } +
e^u ( e^{ u + v } - 1 ) \right]\,. $$
Then, using the boundary conditions (\ref{CFSSEP}) and (\ref{CISSEP}) as
well as (\ref{rotation}), we can evaluate (\ref{act-diff}) : we obtain
\begin{equation}\nonumber
  { 1- r(x) \over 1- \rho(x,0) } { 1-\tilde\rho(x,0)\over 1-1/2} = {
  (e^u-e^v) e^{\tilde H(x,1)} +e^u(e^{u+v}-1) \over (e^u-e^v)
  e^{\tilde H(x,0)} +e^u(e^{u+v}-1)}
\end{equation}
at each $x$, so that $\mu_{\rm an.}(\lambda,\rho_a,\rho_b)=\mu_{\rm an.}(\tilde\lambda,1/2,1/2)$.

\section{Bounds on the decay of the current distribution}
\label{sec:bornes}
In this section, we attempt to generalize the non-Gaussian decay (\ref{tail},\ref{tail-gauss})  of the
distribution of the integrated current $Q_t$ during time $t$,
$$ {\rm Pro}\left[{Q_t\over\sqrt{t}}\simeq q\right]\underset{q\to+\infty}{\asymp} e^{-\alpha\sqrt{t}q^3}\,,$$ 
to other diffusive systems.
We had $\alpha={\pi^2\over 12}$ for the SSEP in the annealed case \cite{DG},
and $\alpha={1\over 12\rho_a^2}$ for non-interacting particles in the quenched
case (\ref{tail-gauss}).

Here, we show that this form of decay holds, both in the annealed and quenched
averages, when the following conditions are satisfied :
\begin{equation}\label{conditions}
  \left\lbrace\begin{array}{lll}D(\rho) &=& 1\,,\\
  \sigma(\rho)& \leq& \rho+c \mbox{ for } 0\leq \rho\leq R \mbox{, with }
  \sigma(\rho)=0 \mbox{ otherwise.}\\
  \end{array}\right.
\end{equation}
More precisely, we set out to show that, when $t\to\infty$ then
\begin{equation}\label{bornes}
  - {q^3\over 2\rho_a\sigma(\rho_a)}\leq{1\over\sqrt{t}}\log {\rm Pro}\left[{Q_t\over\sqrt{t}}\simeq q\right] \leq - {q^3\over 12 (R+c)^2}\,.
\end{equation}
Let 
$$ g(q) = \lim_{t\to\infty} {1\over\sqrt{t}}\log {\rm Pro}\left[{Q_t\over\sqrt{t}}\simeq q\right]\,.$$
In the MFT, $g(q)$ is expressed as the optimum of a variational problem,
like the current generating function $\mu(\lambda)$ (see (\ref{mu_an},\ref{mu_qu})) : 
\begin{equation}\label{g_var}
  g(q) = \max_{\rho(x,\tau),j(x,\tau)} \left\lbrace -{\cal F}_{\rm init.} (\rho(x,0)) -\iint d\tau dx {(j+\partial_x\rho)^2\over 2\sigma(\rho)}\right\rbrace \,,
\end{equation}
where the density profile $\rho(x,t)$ is such that $\int_0^\infty dx \left[
\rho(x,1)-\rho(x,0)\right]=q$, and where the current profile satisfies the
conservation law $\partial_x j+ \partial_t\rho=0$. In addition $\rho(x,0)$ is
free in the annealed case while it is constrained to be equal to $r(x) =
\rho_a + (\rho_b-\rho_a)\theta(x)$ in the quenched case : hence $$g_{\rm
quenched}(q) \leq g_{\rm annealed}(q) \ .$$
Let us first obtain the lower bound in (\ref{bornes}). Because of the
variational formulation (\ref{g_var}) (in the quenched case, ${\cal F}_{\rm
init.}(\rho(x,0))=0$), one can bound $g(q)$ from below by considering a
particular profile $(\rho(x,t),j(x,t))$ leading to a total flux $q$. Here, we
choose to move the segment $[-q/\rho_a,0]$, which contains $q$ particles at
time $0$, at constant speed $v = q/\rho_a$ from time $0$ to time $1$, so that
the total flux through $0$ during this time will be exactly $q$ : this
corresponds to
$$ j(x,\tau) = \left\lbrace \begin{array}{ll}
  q & \mbox{ for }  -q (1-\tau)/\rho_a\leq x \leq q\tau/\rho_a\,;\\
  -\partial_x\rho & \mbox{otherwise.}
\end{array}\right.$$
Since $\rho(x,\tau)=\rho_a$ for $-q (1-\tau)/\rho_a\leq x \leq q\tau/\rho_a$,
this leads to
$$ g(q)\geq -\int_0^1 d\tau \int_{-q(1-\tau)/\rho_a}^{q\tau/\rho_a} 
\hspace{-5mm}dx {q^2\over 2\sigma(\rho_a)} = -{q^3\over 2\rho_a\sigma(\rho_a)}\,,$$
which is the lower bound in (\ref{bornes}) both in the annealed and in the quenched cases.

The upper bound is obtained by noticing that, if $\sigma(\rho)=0$ outside of
$[0,R]$ as in (\ref{conditions}), the fluctuation-dissipation relationship
(\ref{F-eq},\ref{D-sigma}) implies that ${\cal F}_{\rm init.}(\rho(x,0))$
diverges for $\rho(x,0) \notin [0,R]$. From (\ref{conditions}), i.e. 
$\sigma(\rho)\leq \rho+c$, we then obtain
$$g(q) \leq \max_{\rho(x,\tau),j(x,\tau)} - \iint d\tau dx {(j+\partial_x\rho)^2 \over 2(\rho + c)}\,,$$
where $\rho(x,\tau)$ is such that $0 \leq \rho(x,0) \leq R$ and $\int_0^\infty
dx \left[\rho(x,1)-\rho(x,0)\right]=q$. The right-hand side of (\ref{bornesup} 
) is the maximum over $\rho(x,0)$ of the $g_{\rm quenched}(q)$ for
non-interacting walkers with initial density $\rho(x,0)+c$ : it is maximal, 
for $q>0$, when $\rho(x,0)$ is equal to $R$ for $x>0$ and $0$ for $x<0$.
This corresponds to the quenched, non-interactive case (\ref{tail-gauss}) at
densities $R+c$ and $c$, so that $$ g(q) \leq -{q^3\over 12 (R+c)^2}\,,$$
which is the upper bound of (\ref{bornes}).

\section{Conclusion}
\label{sec:conc}
In the present work, we have shown (\ref{conservation-bis}-\ref{mu_qu-bis})
how to implement the macroscopic fluctuation theory to study the fluctuations
of the current of diffusive systems with a step initial density profile. We
have argued that, depending on whether the initial profile can fluctuate or
not, one has to perform an annealed (\ref{mu_an},\ref{mu_an-bis}) or a
quenched average (\ref{t=0-ter}-\ref{mu_qu-bis}).
Using the structure of the equations to be solved in the MFT, we could obtain
a simple relation (\ref{D-sigma-quadratic},\ref{mu-quadratic}) between the
generating functions of the current of the SSEP and of other models with a
quadratic $\sigma(\rho)$ such as the Kipnis-Marchioro-Presutti model. Thus our
solution \cite{DG} for the SSEP determines the generating functions of the
current for all these other models.
We established in section \ref{sec:symmetry} that a time reversal symmetry
(\ref{ft-5},\ref{r(x)-bis},\ref{cond-symmetry}), which is a version of the
fluctuation theorem for a non-steady state initial condition, holds in the
annealed case.
In section \ref{sec:noninter} and in Appendix \ref{gaussien} we showed that
the case of non-interacting particles can be solved both by a macroscopic and
a microscopic approach.
In section \ref{sec:rotation} we have seen that the $\omega$ dependence
of the SSEP could be understood as a rotation invariance of the MFT and we
have exhibited (\ref{rotation}) how the optimal profiles are changed under
these rotations. Lastly, in section \ref{sec:bornes}, we have shown that the
non-Gaussian decay (\ref{tail}) of SSEP is generic under some simple
conditions on $\sigma(\rho)$.

The main difficulty that we could not overcome was to solve the equations
(\ref{conservation-bis}-\ref{t=0-bis},\ref{t=0-ter}) satisfied by the optimal
$\rho(x,\tau)$ and $H(x,\tau)$, even in the case of the SSEP where the
generating function is known. Even for large $\lambda$, we were unable to
solve them, which is why we could only get bounds on the decay of the
distribution of the integrated current $Q_t$ in section \ref{sec:bornes}.
Solving these equations, even in the large $\lambda$ limit, remains an open
question.

\appendix
\section{}
\label{apx:hydro}
In this appendix, we  first show, as in \cite{KTL},
  how  the variational form (\ref{mu_an}) where
one has to optimize over  density  and current profiles  which satisfy
the constraint (\ref{conservation}) can be replaced, using the
Martin-Siggia-Rose formalism,
 by the expression
(\ref{mu_an1}) where the profiles $\rho(x,\tau)$ and $H(x,\tau)$ do not
satisfy any constraint.
We then show that the optimal $\rho(x,\tau)$ and $H(x,\tau)$ are
solutions of (\ref{conservation-bis},\ref{H-evo}) with the boundary
conditions (\ref{t=1-bis},\ref{t=0-bis}).

Let ${\rm Pro}(\rho_0(x)\overset{t}{\rightarrow}\rho_1(x))$ be the probability
of observing the rescaled density profile $\rho_1(x)$ at time $t$, starting
from an initial profile $\rho_0(x)$. Formally, it can be written
(\ref{action}) as a functional integral over all the density and current
profiles  $(\rho(x,\tau), j(x,\tau))$ statisfying $\rho(x,0)=\rho_0(x)$
and $\rho(x,1)=\rho_1(x)$ :
$${\rm Pro}(\rho_0(x)\overset{t}{\rightarrow}\rho_1(x))\asymp \int
{\cal D}\rho{\cal D}j \left[ \prod_{x,\tau}\delta(\partial_\tau \rho +
\partial_x j) \right] 
\exp\left[-\sqrt{t}\iint dx d\tau {(j+D(\rho)\partial_x\rho)^2\over 2\sigma(\rho)}\right]\,,$$
 where the constraint (\ref{conservation})  appears as a $\delta$ function
at each point $(x,\tau)$.
 One can then use an integral representation for each of these $\delta$ functions by introducing   a new field $H(x,\tau)$ :
$${\rm Pro}(\rho_0(x)\overset{t}{\rightarrow}\rho_1(x))\asymp \int {\cal
D}\rho{\cal D}j{\cal D} H \exp\left[-\sqrt{t}\iint dx d\tau
\left( H (\partial_x j + \partial_\tau \rho)  +{(j+D(\rho)\partial_x\rho)^2
\over 2\sigma(\rho)}\right)\right]\,.$$
One can  integrate by parts $\int dx  H\partial_x j$ (this entails
no boundary term as $j$ is expected to vanish at $\pm\infty$) to express
the right-hand side as
\begin{equation}\label{jopt}
  \int{\cal D}\rho
  {\cal D}j{\cal D} H \exp\left[-\sqrt{t}\iint dx d\tau
  \left( H \partial_\tau \rho+D(\rho)\partial_x\rho\partial_x H
  -{\sigma(\rho)\over2} (\partial_x H)^2 +{(j+D(\rho)
  \partial_x\rho-\sigma(\rho)\partial_x H)^2\over 2\sigma(\rho)}\right)
  \right]\,.
\end{equation}
After a Gaussian integration over the currents $j(x,\tau)$
we obtain ${\rm
Pro}(\rho_0(x)\overset{t}{\rightarrow} \rho_1(x))$ as an integral  over the two
unconstrained fields $\rho$ and $H$ :
\begin{equation}\label{action-hrho}
  {\rm Pro}(\rho_0(x)\overset{t}{\rightarrow}\rho_1(x))\asymp \int
  {\cal D}\rho{\cal D} H
  \exp\left[-\iint dxd\tau \left(H \partial_\tau\rho+D(\rho)\partial_x\rho
  \partial_x H  -{\sigma(\rho)\over2}(\partial_x    H   )^2\right) \right]\,.
\end{equation}
Taking (\ref{action-hrho}) together with (\ref{Qt}) and (\ref{pro-init}), one gets 
 $\mu_{\rm annealed}(\lambda)$ as a extremum over
$\rho$ and $H$ : 
\begin{equation}\nonumber
  \mu_{\rm annnealed}(\lambda) = \max_{\rho,H} \left[
  - {\cal F}_{\rm init.}(\rho(x,0))  
  + \lambda \int_0^\infty dx   \; [\rho(x,1) - \rho(x,0) ]  
  -  \iint
  d \tau dx \;\left(H \partial_\tau\rho+ D(\rho)\partial_x\rho \partial_x H-{\sigma(\rho)\over 2}(\partial_x H)^2\right)\right]
\end{equation}
which is (\ref{mu_an1}).
\ \\ \ \\
One can then determine the equations satisfied by the optimal profiles for $\rho$
and  $H$ by looking at the effect of a small variation, 
$\rho(x,\tau) \to \rho(x,\tau) + \delta \rho(x,\tau)$ and $H(x,\tau) \to H(x,\tau) + \delta H(x,\tau)$ : after a few integrations by parts, one
obtains

\begin{eqnarray}\nonumber
  0 &=& \int dx \delta\rho(x,0)\left[-{\delta {\cal F}_{\rm init.}\over
  \delta\rho(x,0)}-\lambda \theta(x)+H(x,0)\right]
  +\int dx  \delta \rho(x,1)\left[\lambda\theta(x)-H(x,1)\right]\\
  & & + \iint d\tau dx \;\delta H(x,\tau)\left[-\partial_\tau\rho+
  \partial_x(D(\rho)\partial_x\rho-\sigma(\rho)\partial_x H)\right]
  \nonumber\\
  & & + \iint d\tau dx \;\delta\rho(x,\tau)\left[\partial_\tau
H  +D(\rho)\partial_x^2  H +{\sigma'(\rho)\over 2}(\partial_x 
 H)^2\right] \label{eqn:var-annealed}\,.
\end{eqnarray}

This yields the two bulk equations (\ref{conservation-bis},\ref{H-evo}) satisfied by $\rho$ and $H$ at the
optimum :
\begin{equation}\nonumber
  \left\lbrace \begin{array}{l}
    \partial_\tau \rho = \partial_x (D(\rho)\partial_x\rho-\sigma(\rho)
    \partial_x H)\\
    \partial_\tau H =-D(\rho)\partial_x^2 H
    -{\sigma'(\rho)\over
    2} (\partial_x H)^2
  \end{array}\right.\,.
\end{equation}

The first of these equations is just the  conservation law,
$\partial_x j + \partial_x \rho =0$, since, from (\ref{jopt}), we have
$$j=-D(\rho )\partial_x\rho+\sigma(\rho)\partial_x H $$
at the optimum. Using (\ref{F-init}) to express ${\delta {\cal F}_{\rm 
init.}\over\delta\rho(x,0)}$, we also obtain from (\ref{eqn:var-annealed}) the boundary relationships
\begin{equation}\nonumber
  \left\lbrace \begin{array}{l}
  H(x,1) = \lambda\theta(x)\\
  H(x,0) = \lambda\theta(x) + f'(\rho(x,0))-f'(r(x))
  \end{array}\right.\,,
\end{equation}
which reduce to (\ref{t=1-bis},\ref{t=0-bis}) by using (\ref{D-sigma}).

\section{}
\label{apx:sym}

In this appendix, we show that   when $D(\rho)$ and $\sigma(\rho)$
satisfy the particle-hole symmetry  (\ref{D-sigma-sym}), the optimal profile (assuming that it is unique) in (\ref{mu_an}) verifies (\ref{part-hole-sym}) when $\rho_a=\rho_b=\bar\rho$. This will allow us to  relate the optimal profiles in the annealed and in the quenched cases and to obtain (\ref{part-hole-sym1}). 

First, when $\rho_a=\rho_b (=\bar\rho)$, the term proportional to $f'(r(x))$ in (\ref{ft-4}) vanishes due to the conservation of the total number of particles, so that (\ref{ft-4}) becomes
\begin{eqnarray}
\mu_{\rm annnealed}(\lambda) = \max_{\rho(x,\tau),j(x,\tau)} \left\{-
  { {\cal F}_{\rm init.}(\rho(x,1))
 +{\cal F}_{\rm init.}(\rho(x,0)) \over 2}
+  \int_{-\infty} ^\infty dx   \; [\rho(x,1) - \rho(x,0) ]  \lambda \theta(x)  
 \right.  \nonumber  \\  \left. -  
\int_{0}^{1}
d \tau \int_{-\infty}^\infty dx \;  { j(x,\tau)^2 + (D(\rho(x,\tau))
 \partial_x\rho(x,\tau))^2 \over 2
\sigma( \rho(x,\tau))} \right\}\,.  &&
\label{ft-7}
\end{eqnarray}
Moreover (\ref{D-sigma-sym}) implies (see(\ref{F-init-1})) that
$${\cal F}_{\rm init.}(\rho(x,\tau)) =   {\cal F}_{\rm init.}(2 \bar\rho -\rho(x,\tau))\,.$$
Therefore, if $(\rho^{(a)}(x,\tau),j^{(a)}(x,\tau))$ is optimal in (\ref{ft-7}), then $(2 \bar\rho-\rho^{(a)}(x,1-\tau),j^{(a)}(x,1-\tau))$
is also optimal
and, if this optimum is unique, one gets (\ref{part-hole-sym})
\begin{equation}
\rho^{\rm (a)}(x,\tau) =  2 \bar \rho - \rho^{\rm (a)}(x,1-\tau)  \ .
\label{ph}
\end{equation}
Due to this symmetry, one can rewrite (\ref{ft-7}) as
\begin{eqnarray}
\mu_{\rm annnealed}(\lambda) = 2 \max_{\rho(x,\tau),j(x,\tau)} \left\{
  { -{\cal F}_{\rm init.}(\rho(x,1))
  \over 2}
+  \int_{-\infty} ^\infty dx   \; [\rho(x,1) - \bar\rho ]  \lambda \theta(x)  
 \right.  \nonumber  \\  \left. -  
\int_{1/2}^{1}
d \tau \int_{-\infty}^\infty dx \;  { j(x,\tau)^2 + (D(\rho(x,\tau))
 \partial_x\rho(x,\tau))^2 \over 2
\sigma( \rho(x,\tau))} \right\}\,.  &&
\label{ft-8}
\end{eqnarray}
with $\rho^{\rm (a)}(x,\tau=1/2)= \bar\rho$ from (\ref{ph}).

For the quenched problem, using the identity (\ref{identity}), the fact that
the term proportional to $f'(r(x))$ vanishes, and that ${\cal F}_{\rm
init.}(\rho(x,0)=\bar\rho)=0$, one can rewrite (\ref{mu_qu}) as
\begin{eqnarray}
\mu_{\rm quenched}(\lambda) = \max_{\rho(x,\tau),j(x,\tau)} \left\{
  { -{\cal F}_{\rm init.}(\rho(x,1))
  \over 2}
+  \int_{-\infty} ^\infty dx   \; [\rho(x,1) - \rho(x,0) ]  \lambda \theta(x)  
 \right.  \nonumber  \\  \left. -  
\int_{0}^{1}
d \tau \int_{-\infty}^\infty dx \;  { j(x,\tau)^2 + (D(\rho(x,\tau))
 \partial_x\rho(x,\tau))^2 \over 2
\sigma( \rho(x,\tau))} \right\}\,,  &&
\label{ft-9}
\end{eqnarray}
with the initial-time condition $\rho(x,0)=\bar\rho$.

We see that (\ref{ft-8}) and (\ref{ft-9}) are identical except for the range of variation of $\tau$.
This allows us to relate the optimal profiles in the annealed and the quenched cases by
\begin{equation}\nonumber
  \left\lbrace \begin{array}{lll}
    \rho^{\rm (q)}(x,\tau) &=&  \rho^{\rm
(a)}\left({x\over\sqrt{2}},{1+\tau\over2}\right)\\
    j^{\rm (q)}(x,\tau) &=& {1\over\sqrt{2}} j^{\rm (a)}
\left({x\over\sqrt{2}},{1+\tau\over2}\right)
  \end{array}\right.\,,
\end{equation}
from which (\ref{part-hole-sym1}) follows easily.

\section{}
\label{gaussien}

In this appendix, we first show why, for non interacting walkers on a one
dimensional lattice as in section \ref{sec:noninter}, $D(\rho),\sigma(\rho)$
and $f(\rho)$ are given by (\ref{D-sigma-f-gaussien}). We then explain how
(\ref{mu_an-ter3}) and (\ref{mu_qu-ter2}) can be recovered by a microscopic
calculation.

Consider first a 1d lattice of length $L$ : a new particle is injected at rate
$\alpha$ on site $1$ and at rate $\delta$ on site $L$. Each particle on site
$1$ is removed at rate $\gamma$ and on site $L$ at rate $\delta$. As the
particles do not interact, the probability that a particle $T_i$ on site $i$
will have escaped, after time $t$, into the right reservoir evolves according
to
\begin{eqnarray*}
{d T_1 \over dt}&=& T_2 - (1+\gamma) T_1  \,;\\
  {d T_i \over dt}&=&  T_{i+1} + T_{i-1}  -  2 T_i \qquad \mbox{for } 2\leq i \leq L-1\,;\\
  {d T_L \over dt}&=&   \beta + T_{L-1} -(1+\beta) T_L\,,
\end{eqnarray*}
whose solution in the long time limit is
$$T_i = {i  + {1 \over \gamma} - 1  \over L-1 + {1 \over \beta} + {1 \over \gamma}}\,.
$$
It is easy to see that the contribution to $Q_t$ of the particles entering the system during the first time interval $dt$ is
$$\langle e^{\lambda Q_{t+dt}} \rangle =
\langle e^{\lambda Q_{t}} \rangle  \left( 1 + \alpha  T_1 ( e^\lambda -1) dt + \delta (1-T_L)(e^{-\lambda} -1) dt \right) \,.
$$
Therefore
$$\lim_{t \to \infty} {1 \over t} \log \langle  e^{\lambda Q_t } \rangle =
{ {\alpha \over \gamma}(e^{\lambda } -1)
+ {\delta \over \beta}(e^{-\lambda } -1)
\over
L-1 + {1 \over \beta} + {1 \over \gamma}}\,,
$$
which becomes for large $L$ 
$$\lim_{t \to \infty} {1 \over t} \log \langle  e^{\lambda Q_t } \rangle =
{  \rho_a(e^{\lambda } -1)
+ \rho_b(e^{-\lambda } -1)
\over
L} \,,
$$
with $\rho_a= {\alpha \over \gamma} $ and $\rho_b = {\delta \over \beta}$.
The expansion in powers of $\lambda$ (see (\ref{D-def},\ref{sigma-def})) leads
to $D(\rho)=1$ and $\sigma(\rho)= 2 \rho$, as in (\ref{D-sigma-f-gaussien}).
For these non interacting particles, the partition function $Z(N,L)=L^N/N!$ so
that $$f(\rho)= \rho \log \rho - \rho$$ as in (\ref{D-sigma-f-gaussien}) , and
(\ref{D-sigma}) is verified.

One can also see that at equilibrium, at density $\rho$, there is an invariant
measure (the equilibrium) where the occupation numbers $n_i$ of the sites are
independent random variables distributed according to a Poisson distribution
\begin{equation}\label{Poisson}
  {\rm Pro}(n) =  {Z(N-n,L-1) Z(n,1) \over Z(N,L)}  \simeq {\rho^n  \; e^{- \rho} \over n!} \,. 
\end{equation}

Let us now consider non-interacting particles on an infinite one dimensional
lattice. Each particle jumps at rate 1 to each of its neighboring sites. The
probability $P_{i,j}(t)$ that a particle initially at position $i$ will travel
a distance $j-i$ is given, for large $t$, by
\begin{equation}
P_{i,j}(t)  \simeq {1 \over \sqrt{4 \pi t}} e^{-{(j-i)^2 \over 4 t}}\,.
\label{Pij}
\end{equation}
The contribution of a particle initially located at site $i$ to $e^{\lambda
Q_t}$ is
$$ \Phi_{i}=1 + (1-\theta_i) (e^{\lambda} -1) \sum_ {j \geq 1} P_{i,j}(t) +
\theta_i (e^{-\lambda} -1) \sum_ {j \leq 0} P_{i,j}(t) \,,$$
where $\theta_i=1$ if $i \geq 1$ and $\theta_i=0$ if $i \leq 0$. In the long
time limit, this becomes
$$ \Phi_{i}=1 + (1-\theta_i) (e^{\lambda} -1) {1 + E({i \over 2 \sqrt{t}})
\over 2} + \theta_i (e^{-\lambda} -1) {1 - E({i \over 2 \sqrt{t}}) \over 2}
\,, $$
where $E$ is the error function defined in (\ref{error}).
Therefore, for a given initial condition where the occupation numbers $n_i$ of
all the sites are specified, one gets
$$ \left\langle e^{\lambda Q_t} \right\rangle_{\rm history} = \exp\left[
\sum_i n_i \log \Phi_i \right] \,.$$

The $n_i$ are distributed according to a Poisson distribution (\ref{Poisson})
of density $\rho_a$ on the negative axis and $\rho_b$ on the positive axis.
Averaging over the $n_i$ (i.e. over the initial conditions) leads to
(\ref{mu_an-ter3}) in the annealed case and to (\ref{mu_qu-ter2}) in the
quenched case.



\end{document}